\documentclass[a4paper,11pt]{article}
\usepackage{pos}

\usepackage{wrapfig}

\usepackage[rightcaption]{sidecap}

\newcommand{\lhuc}{LH\ensuremath{\mu}C}
\newcommand{\lhmuc}{LH\ensuremath{\mu}C}

\title{LH$\mu$C: A high energy, high luminosity muon-hadron collider at CERN}

\author[a]{K. Piotrzkowski}

\onbehalf{on behalf of the HELP4CERN Study Group$^\dagger$}

\affiliation[a]{AGH University of Krak\'ow, Faculty of Physics and Applied Computer Science, Al. Mickiewicza 30, 30--055 Kraków, Poland}

\emailAdd{piotrzkowski@agh.edu.pl}

\abstract{The latest developments regarding the feasibility of muon-hadron collisions, achieved through the major reuse of LH{\it e}C infrastructure, are presented -- at the \lhmuc, a 1 or 2 TeV $\mu^+$ beam collides with a 7 TeV proton beam, resulting in the $\mu^+p$ centre-of-mass energy of 5.3 or 7.5~TeV, respectively, at an average luminosity close to $10^{34}$cm$^{-2}$s$^{-1}$. First, a concise scientific motivation is given, including  exceptional prospects for neutrino science, and is followed by a presentation of the  LH$\mu$C concept in a nutshell -- from the source of ultra-cold anti-muons, through the muon booster, to the main muon ring. Basic considerations regarding the machine-detector interface and the general-purpose detector are also presented. The paper concludes with an outlook underlying unique and powerful synergies between the LH$\mu$C, LH{\it e}C, FCC and Muon Collider projects.\\
\vspace{1cm}

{\footnotesize $^\dagger$ {\bf H}igh {\bf E}nergy {\bf L}epton {\bf P}roton collisions {\bf for CERN} Study Group: \it D. Acosta, D. Britzger, A.C.~Canbay, K.~Cheung, B. Dagli, L. Forthomme,
B. Holzer, U. Kaya, H. Khanpour, A. Kilic, F. Kocak, P.~Kotko, M.~Kumar, K. Long, X. Lu, B.~Mellado, S. Moretti,
M. Otani, K. Piotrzkowski, C. Rogers, D. Schulte, A.M.~Sta\'sto, S.~Sultansoy, I. Tapan,
S. Tentori, M.~Topp-Mugglestone, Z.S. Wang, Y. Yamazaki, C. Zhang, G.~Ziarko, F. Zimmermann.}

{\hspace{4.5cm}{\includegraphics[height=22mm]{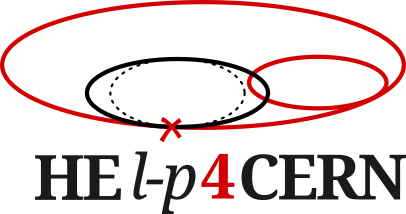}}}}

\FullConference{The 33rd International Workshop on Deep Inelastic Scattering and Related Subjects (DIS2026)\\
4 - 8 May 2026\\
Bologna, Italy\\}

\begin{document}
\maketitle

\section{Introduction}
\noindent
This paper provides a brief report on the current status of the work being conducted by the HELP4CERN Study Group~\cite{white-paper}, triggered by the recent $\mu$TRISTAN proposal~\cite{Hamada:2022mua}, which in turn was utilized to consider muon-hadron collisions at the HL-LHC~\cite{Akturk:2025ubm}, using the energy recovery linac infrastructure of the electron-hadron collider LH$e$C~\cite{LHeCStudyGroup:2012zhm,LHeC:2020van,Ahmadova:2025vzd}. Contributions to the first \lhmuc\ workshop offer more insight into these developments~\cite{ultra}.

The collisions of 1~TeV muons with 7 TeV protons at \lhmuc, corresponding to $\sqrt{s_{\mu p}}=5.3$~TeV (therefore significantly higher than the $ep$ centre-of-mass energy at the proposed FCC-$eh$) and  the peak \lhmuc\ luminosity which might exceed $10^{34}$cm$^{-2}$s$^{-1}$, are effectively making it a powerful Higgs-Electroweak-Top (HET) factory. Moreover, an option of the 2 TeV muon beam is also considered, resulting in a yet higher $\sqrt{s_{\mu p}}=7.5$~TeV.

\begin{SCfigure}[0.36][h]
\centering
\includegraphics[width=.7\textwidth]{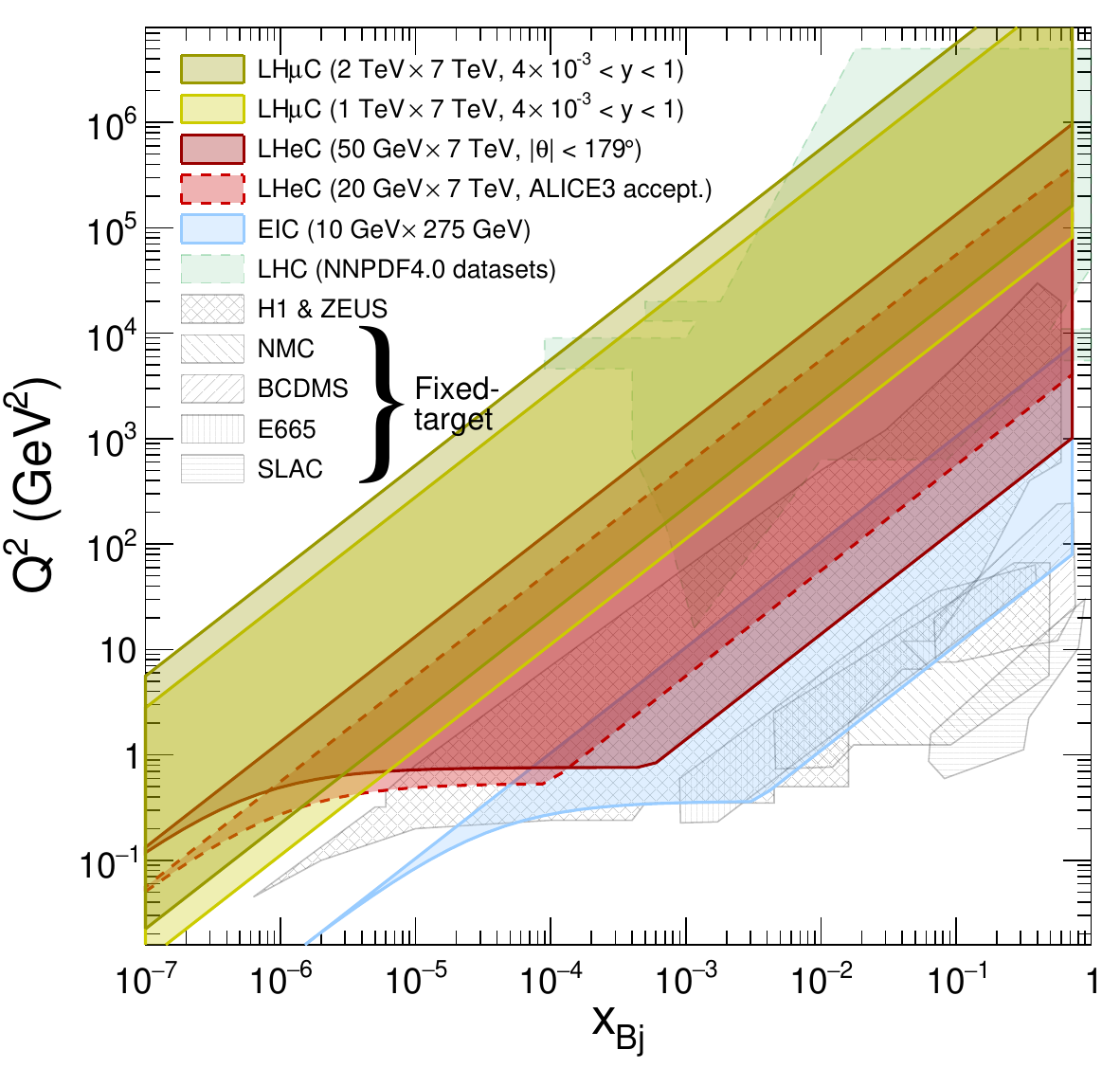}
\vspace{-3mm}
\caption{Kinematical reach on the $(x_{Bj},Q^2)$ plane for the lepton-proton collisions at the LH$e$C and \lhuc, in comparison to existing fixed target and collider datasets, and planned EIC experiments.\\\\}
\label{fig:xbj_vs_q2_phase_space}
\end{SCfigure}

Already an initial running of \lhmuc\ at very low luminosity will allow for a huge kinematical expansion in the $(x_{Bj}, Q^2)$ plane for studying QCD: by more than one order of magnitude with respect to the LH$e$C and by more than two orders wrt. HERA (see Fig.~\ref{fig:xbj_vs_q2_phase_space}). 
As a result, unique studies of parton non-linear effects at the LH$e$C can be further extended~\cite{piotr}, including powerful insights into elusive nuclear effects in the lepton-ion scattering~\cite{LHeC:2020van,Ahmadova:2025vzd,burak}. In addition, PDF flavor decomposition can also be further improved, as well as precision measurements of $\alpha_S$, which are of high relevance for research at the HL-LHC and future hadron colliders.

In the context of the determination of electroweak parameters, \lhuc\ also offers improvements over the LH$e$C~\cite{white-paper,Ahmadova:2025vzd,hamzeh} by precision measurements of $M_W$ and $\sin^2{\theta_W}$, due to the significantly larger span in $Q^2$. In addition, LH$\mu$C will become a top-quark factory, thanks to the high luminosity and high $\sqrt{s_{\mu p}}$, and will be particularly well suited for precision studies near the production threshold and for unique measurements of the top charge~\cite{white-paper, hamzeh}. Moreover, similarly to the LH$e$C, it is expected to make significant contributions to studies of single top production and rare top decays. Furthermore, due to its high centre-of-mass energy and unique initial state, \lhuc\ will also serve as a powerful platform for BSM searches~\cite{Akturk:2025ubm,Cheung:2021iev}, and $\gamma\gamma$ physics~\cite{hamzeh2}.

In the Higgs boson sector, \lhuc\ offers similar, or even better, sensitivity to the Higgs boson couplings than LH$e$C~\cite{white-paper,simone}. Particularly valuable will be the \lhuc\ contributions to the precise determination of the Higgs couplings to Z and W bosons, which will remain highly relevant even during the FCC-$ee$ era~\cite{Ahmadova:2025vzd,Cheung:2021iev,simone}. Last but not least, thanks to its very large centre-of-mass energy, the \lhmuc\ might offer a possibility of measuring the Higgs boson self-couplings -- studies of such measurements were performed in the context of the FCC-$eh$ proposal and provided promising results~\cite{Kumar:2015kca}. The preliminary results obtained for the \lhmuc\ case are equally exciting~\cite{mukesh}.

Finally, as a free bonus, the \lhmuc\ complex will provide very powerful neutrino beams, offering in Europe a potent and unique research platform for neutrino science.
\section{\lhmuc\ Concept} 
\noindent
Following the $\mu$TRISTAN proposal~\cite{Hamada:2022mua}, the LH$\mu$C concept primarily relies on the source of ultra-cold anti-muons recently developed at J-PARC and KEK~\cite{Aritome:2024rlu}, rather than using the conventional ionisation cooling approach. Although this method also begins with a proton driver -- Rapid Cycling Synchrotron (RCS) -- and a pion production target (see Fig.~\ref{subsec:LHuC}), the muon cooling goes via stopping muons from the pion decays in aerogel where "thermal" muonia are formed and then, after diffusing out, are ionised by a laser. Clearly, this technique is only applicable for the $\mu^+$ production.

\label{subsec:LHuC}
\begin{figure}[h]
\centering
\includegraphics[width=1.0\textwidth]{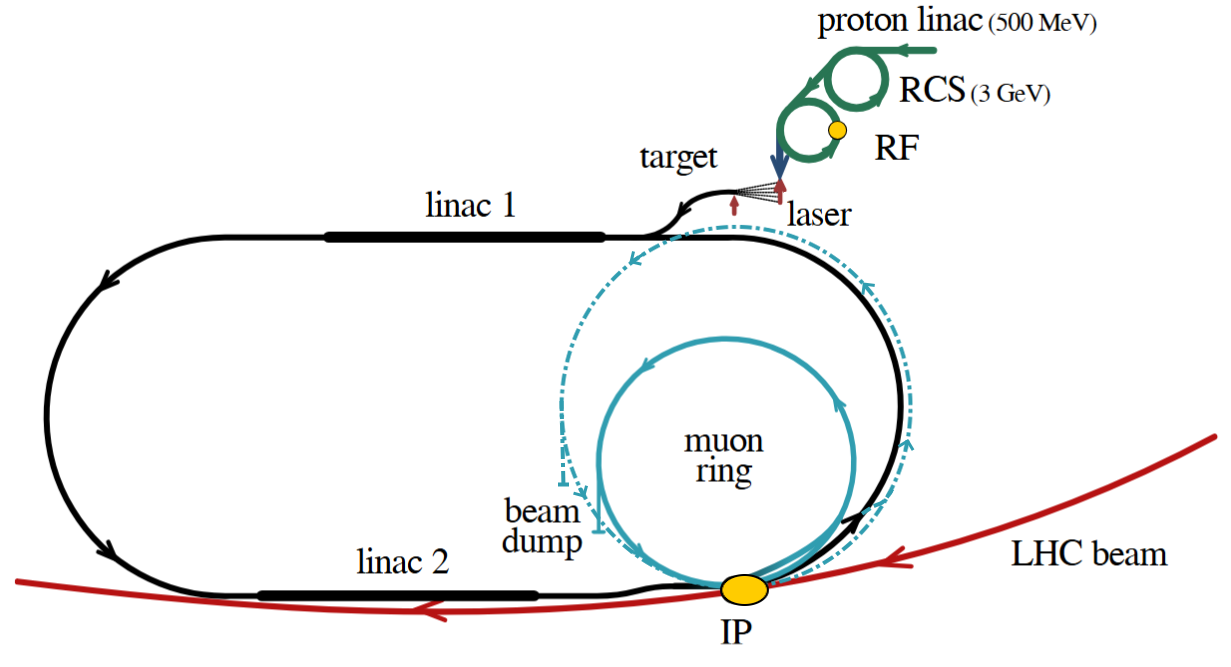}
\caption{Layout of the \lhuc~concept, based on the $\mu$TRISTAN proposal~\cite{Hamada:2022mua}. The ERL components for the lepton beam acceleration are depicted in black, while the 1~TeV muon storage ring is shown in light blue. The dash-dotted circle represents the 2 TeV muon ring. \label{fig:i}}
\end{figure}

To achieve high yield of ultra-cold anti-muons, two major modifications were proposed: re-use of the proton beam for multiple target hits (hence the need for the second RCS) and a different aerogel configuration~\cite{Hamada:2022mua,cedric}. Very recently, in new studies hot tungsten wires were considered in place of aerogel slabs, and the cooling process was proposed to be carried out in two steps~\cite{Sakaki:2026uvl}. Moreover, two-step cooling is an essential part of a new hybrid approach to an ultra-cold $\mu^+$-source~\cite{chris}. Last but not least, the method of pre-acceleration of ultra-cold anti-muons to a GeV energy is also actively pursued~\cite{Aritome:2024rlu,masashi}.

The rapid acceleration of anti-muons from an energy of 1~GeV  to the collision energy of 1 or 2~TeV is one of the most critical challenges of the \lhuc\ design. 
To circumvent the fundamental limitations of fast-ramping magnetic fields inherent in RCS designs, a novel acceleration scheme based on the vertical Fixed-Field Alternating Gradient (vFFAG) principle~\cite{max} is proposed. For example, if one assumes a three-pass acceleration to 70~GeV, inherited from the nominal LH$e$C design, adding one vFFAG beamline might allow for further acceleration up to 500~GeV~\cite{max2}. Therefore, one can expect to reach a 1~TeV energy and beyond using a one-pass LH$e$C~\cite{Andre:2025iar} and two vFFAG beamlines, for example. The 1(2)~TeV anti-muon beams emerging from the booster of about 9~km circumference will be transferred to the circular main ring of about 3(6)~km positioned tangentially to the LHC (see Fig.~\ref{subsec:LHuC}). Finally, it should be noted that high polarisation of the $\mu^+$ beam might also be feasible~\cite{Hamada:2022mua}.

\begin{wraptable}{r}{0.52\textwidth}
\vspace{-2mm}
\centering
\caption{Example of the LH$\mu$C parameters.}
\label{tab:final_performance}
\begin{tabular}{lcc}
\hline
\textbf{Parameter} & \multicolumn{2}{c}{\textbf{Value}} \\
\hline
Center-of-mass energy, $\sqrt{s}$ [TeV] & \multicolumn{2}{c}{5.3} \\
Peak luminosity, $\mathcal{L}$ [$10^{34} \text{ cm}^{-2}\text{s}^{-1}$] & \multicolumn{2}{c}{2.1} \\
Collision({Repet.}) rate [MHz(Hz)]&\multicolumn{2}{c}{4(12.5)} \\
\hline
\textbf{Beam parameters} & $\mu^+$ & $p$ \\
\hline
RMS beam size at IP, $\sigma_{x,y}$ [$\mu$m] & 4.8 & 4.8 \\
Bunch population [$10^{11}$] & 0.07 & 22\\
\hline
\end{tabular}
\end{wraptable} 
Assuming rather conservatively the $\mu^+$ yields at a level of 5\% of the ones in the $\mu$TRISTAN 
  proposal, a 4~MW proton driver, and the possibility of an increased population of LHC proton bunches due to their smaller number  (hence, lower collision rate)~\cite{frank}, the performance of \lhmuc \ is very compelling, as shown alongside.
The $\mu^+$ train stored in the main ring in this case contains 80 bunches separated by 250~ns, and the repetition rate is 12.5~Hz.

Last but not least, the \lhmuc\ complex will produce, as a "side effect",  $\nu_e$ and $\bar{\nu}_\mu$ beams -- very intense and highly collimated in three different directions -- along two booster linacs as well as the straight section of the main ring. The energies of the decaying anti-muons are distributed in complementary bands, from 1~GeV to the top energy.

\section{Interaction Region}
\noindent
In general, there are two options to consider for the focusing of muon beam: either one makes focusing of muons close to the IP, "in between" the mini-beta quadrupoles of the LHC, as is proposed for electrons at the LH$e$C; or focusing of muons is "outside" mini-beta quadrupoles of the LHC,
in a "linear collider type style".
The time separation of muon bunches is large at \lhuc, therefore, a very small crossing angle is needed to avoid parasitic collisions of muon and proton bunches. To achieve a "physical" separation of the muon and proton beams, the most preferred scheme is by introducing a new separator dipole, and
one should note in that context that the problem of making high energy $\mu^+p$ collisions is in general very similar to the $pp$ case at the LHC.
Last but not least, the appropriate shielding of the beam-lines against the background due to the beam-muon decays has to be studied and properly designed, but one can profit here from the corresponding in-depth studies performed for the Muon Collider project.  

\section{General Purpose Detector}
\noindent
The proposed main LH$e$C detector~\cite{Ahmadova:2025vzd,Andre:2022xeh} has many of the key features required for high-precision muon-hadron measurements at \lhmuc, including hermetic coverage and well-balanced tracking and calorimetry systems. Thus, most segments of the baseline LH$e$C detector can be expected to remain directly applicable after only relatively minor modifications~\cite{laurent}. 
However, in place of the electron beam-induced backgrounds at the LH$e$C due to strong synchrotron radiation, at \lhmuc\ one has to deal with very different backgrounds due to the beam-muon decays. 
It requires an optimal design of dedicated shielding: the so-called $nozzle$ should be located inside the \lhmuc\ detector in the direction of incoming muons~\cite{nozzle}, its size has to be carefully optimised to balance the loss of instrumentation in the volume it occupies and the efficiency of the background suppression.

In most of the interactions at \lhmuc\ one expects a TeV-muon to be present in the final state. Very initial considerations indicate that for high $p_T$ muons, a CMS-like muon detector might be the optimal choice in that context~\cite{yuji}. On the other hand, the detection of incident muons scattered at small angles will require special solutions~\cite{darin}.
However, one should not forget here that there are alternative ways to properly reconstruct event kinematics in $\mu p$ interactions, without a precise measurement of the energy of the scattered muon, but using the measured momentum of "the recoil" jet and  muon scattering angle~\cite{Acosta:2022ejc}.

Precise measurements of the absolute \lhmuc\ luminosity are crucial for its scientific programme -- at HERA the electron-proton bremsstrahlung was successfully used for that purpose, and this is proposed for the EIC, also for the electron-ion case~\cite{Piotrzkowski:2021cwj}. The first considerations indicate that muon-hadron bremsstrahlung should provide the necessary means~\cite{krzysztof}.

It should be noted that the event pileup at \lhmuc\ will be negligible or small, depending on the actual bunch populations. Therefore, not only the proper detection and reconstruction of complex final states will be possible but also application of "triggerless" data-streaming systems for the data acquisition, as is already planned for the ALICE3 and ePIC experiments, for example.

\section{Outlook}
\noindent
The realisation of the \lhmuc\ proposal will provide a versatile and profound extension of the scientific programme at the HL-LHC, in addition to offering "for free" a unique and powerful platform at CERN for neutrino science.
The \lhuc\ project, on the one hand, profusely benefits from the developments of the Muon Collider concept and of cold muon beams in general, but on the other hand it would constitute a major "test-bench" for these projects. Furthermore, in the long term, the \lhuc\ project potentially opens amazing opportunities to greatly extend the scientific scope of the proposed FCC-$hh$ (or HE-LHC) facility by adding, at a fractional cost, a very powerful muon-hadron collider, capable of colliding muons and protons at 30 (15)~TeV, and in addition, of producing very high flux neutrino beams. 
Finally, in the spirit of the original $\mu$TRISTAN proposal, a very exciting possibility of making high energy, high luminosity $\mu^+e^-$ collisions at the FCC-$ee$ (or at LEP3) is also being considered~\cite{krzysztof2}.
\section*{Acknowledgements}
{
\noindent
K.P. is supported by the NCN grant 2023/49/B/ST2/03273 \& NAWA grant BPN/PPO/2021/1/00011.}
\small
\bibliographystyle{JHEP}
\bibliography{biblio}

\end{document}